\documentstyle[aps,prb]{revtex}

\begin{document}
\twocolumn[\hsize\textwidth\columnwidth\hsize\csname@twocolumnfalse\endcsname

\title{Plasmonic excitations in noble metals: The case of Ag.}
\author{M.~A. Cazalilla, J.~S. Dolado}
\address{Materialen Fisika Saila,
 Euskal Herriko Unibertsitatea,
1072 PostaKutxatila, 20080 Donostia, Spain}
\author{A. Rubio}
\address{Departamento de F\'{\i}sica Te\'orica,
Universidad de Valladolid,
E-47011 Valladolid, Spain}
\author{P.~M. Echenique}
\address{Materialen Fisika Saila,
 Euskal Herriko Unibertsitatea,
1072 Postakutxatila, 20080 Donostia, Spain\\
and Centro Mixto CSIC-UPV/EHU, Spain\\}

\date{\today}

\maketitle

\begin{abstract}
The delicate interplay between plasmonic excitations and
interband transitions in noble metals is described by means of
{\it ab initio} calculations and a simple model in which 
the conduction electron plasmon is coupled to the 
continuum of  electron-hole pairs. 
Band structure effects, specially the energy at which the excitation 
of the $d$-like bands takes place, determine the existence of a subthreshold plasmonic mode,
which manifests itself in Ag as a sharp resonance at $3.8$ eV. 
However, such a resonance is not observed in the other noble metals.
Here, this different behavior is also analyzed and 
an explanation is provided.
\end{abstract}

\pacs{PACS numbers: }
]

\section{Introduction}\label{intro}

In a recent letter,~\cite{KE99} Ku and Eguiluz have shown that
band structure effects are responsible for the positive 
dispersion of the plasmon line width in K. Such a behavior had been
previously found by vom Felde, Spr\"osser-Prou, and
Fink~\cite{FSF89} in electron energy loss experiments. They
interpreted the result as due to short range electron-electron
correlations beyond the random phase approximation (RPA). However,
Ku and Eguiluz explain the result within the RPA in terms of the
extra plasmon decay channels provided by a manifold of unoccupied
d-symmetry bands.  The effect of these localized bands is
overlooked when treating K as an homogeneous electron system in
the {\it jellium} model. Therefore, band structure effects turn 
out to be decisive in determining, not only the 
existence interband transitions, but also  the
properties of plasmonic excitations. Indeed, {\it ab initio} 
calculations of the dynamical response  have
properly described the experimental plasmon dispersion in a
variety of simple metals.~\cite{QE93,AK94,FSE97,MG94}
On the other hand, the {\it jellium} model 
still provides a qualitative, though
not accurate, description of the  energy loss spectrum of
these systems in the long wavelength limit. 

 The energy loss spectrum can be obtained either by doing electron
energy loss spectroscopy~\cite{R80} (EELS), inelastic scattering of X-ray
photons~\cite{SSSK95}(IXSS), or indirectly by means of optical 
measurements.~\cite{W72} In noble metals this spectrum has a 
complicate structure, which bears little resemblance
to that of simple metals. It is known~\cite{EP62} 
that such a structure has to do with the existence of a manifold of 
$d$-like bands a few eV below the Fermi level. These lead to important 
deviations from even the crudest predictions for an
homogeneous electron liquid. Thus, whereas for Ag a narrow
resonance is observed at an energy $\approx 3.8$ eV, no
well-defined plasmonic excitations appear in the spectra of Cu and Au.

 The purpose of this paper is to investigate
the structure of the energy loss spectrum in the long wavelength
limit, where plasmonic excitations are expected to be important. 
In particular, we have focused on Ag for our study, but
some of our conclusions do also apply to Cu and Au, 
and could be extended to similar metallic compounds with
fully occupied $d$-like bands close to the Fermi level. The spectrum
of Cu was studied from first principles in a previous
work.~\cite{CRP99} In the present work, 
we show that the structure of the energy loss spectrum
of Ag up to $\sim 10$ eV can be understood in terms of a 
simple model. In particular, we are concerned with the role played by
the relative values of the Drude plasma frequency and the
threshold energy for the excitation of the $d$-like bands. 
After clarifying the relevant elements for the existence
of a plasmonic excitation below the interband excitation threshold,
numerical experiments in an {\it ab initio} framework 
for calculating the dielectric matrix of Ag
are performed to confirm this picture. Furthermore, 
comparison  of the {\it ab initio} results with experimental data
gives good agreement for excitations well above threshold,
being able to reproduce all the structure in the loss 
spectrum up to $\approx 30$ eV. The underestimate 
of the threshold energy in local density 
calculations also shows up in our calculations, 
giving a lower plasma frequency than the experiments
(see below for a more detailed discussion). 

 The question of how band structure alters the properties
of plasmons in metals has been previously addressed by a number of
authors. Wilson~\cite{W60} has studied the shift in the Drude plasma
frequency by the presence of an optical absorption band that lies
above (below) the free electron plasma frequency. He has found
that the plasma frequency is shifted
downwards (upwards) by the effect of the absorption band. More recently, 
Sturm~\cite{S82} has studied within the RPA the corrections to the dielectric
function of an homogeneous electron gas due to a weak lattice
potential. He has shown that the lattice potential can provide the
necessary momentum  for a plasmon to decay into an electron-hole
pair below the Landau cutoff.~\cite{landau_cutoff} In other words,
when taking into account the lattice structure by folding the free
electron bands the first Brillouin zone (BZ), a
plasmonic excitation can decay by producing an interband
transition. As a consequence, even in th RPA, plasmons 
acquire a finite lifetime. Core polarization effects on the plasma frequency
have been investigated by Zaremba and Sturm.~\cite{ZS85} They treated the
core polarizability as the superposition of the polarizability of the
isolated lattice ions. The polarization of core electrons lowers the plasma
frequency from its Drude value,  bringing it closer to the experimental value.

 Recently, the dielectric matrix as well as the energy loss 
spectrum have been evaluated numerically within a first 
principles framework. In these calculations, one usually starts from a set 
of  LDA~\cite{KS65} orbitals and energies, 
using time-dependent density functional theory (TDDFT),~\cite{tddft} 
an approximation to the density correlation function 
can be calculated. Hence, The energy loss spectrum is obtained
from the imaginary  part of the density correlation function, as dictated by
the fluctuation-dissipation theorem.~\cite{PN66}
To mention a few examples of theses studies, besides
the above mentioned work by Ku and Eguiluz~\cite{KE99},
Quong and Eguiluz~\cite{QE93} have thus investigated the 
anisotropy in the plasmon dispersion of
aluminum~\cite{UR76}. Aryasetiawan and Karlsson~\cite{AK94} have
studied the excitations in the energy loss spectra of Li. The
negative dispersion of bulk plasmons in Cs has been explained by
Fleszar and co-workers.~\cite{FSE97} In general, specific details
of the band structure of metals as the presence of band gaps in
the conduction band region are needed to describe the experimental
energy loss spectra in some simple~\cite{MG94,F95} and noble 
metals.~\cite{CRP99}

  The behavior of noble metals has been considered by 
Ehrenreich and Phillipp.~\cite{EP62} These authors used the data
obtained  in experimental measurements of the reflectivity from Cu
and Ag~\cite{TP61} to obtain the optical and energy loss spectra. 
They separated the Drude and
interband contributions to the optical response, thus shedding
light on the reason why the energy of the plasma resonance
in Ag is shifted down in energy. In a later work, Cooper,
Ehrenreich and Phillipp~\cite{CEP65} extended this analysis to Au.
They touched upon the question of assigning interband
transitions to the most relevant features observed in the optical
spectrum. To further clarify this point, Mueller and
Phillips~\cite{MP67} performed,  within the RPA and based on 
band structure calculations  by  Burdick,~\cite{B63} a numerical
calculation of the imaginary part of interband contribution to the
optical response of Cu. Thus, they could give a more 
correct identification of the interband transitions
that contribute to the optical spectrum at a given energy. Their
interpretation is consistent with the optical data obtained in a
number of reflectivity measurements from Ag and Ag alloys.~\cite{ML68,IHW70}

 As we have mentioned above, the energy loss function of Cu has been
recently  evaluated from first principles by Campillo, Rubio, and
Pitarke.~\cite{CRP99} These authors solved the
Kohn-Sham~\cite{KS65} equations on a plane wave basis using a norm
conserving pseudopotential.~\cite{TM91} Subsequently, they evaluated the
dielectric and energy loss functions. The method is shown to 
give an accurate description of the energy loss spectrum of Cu once 
the fully occupied $d$-like bands are taken into
account as part of the valence electron complex.
The remaining core electrons
do not take an active part in the excitations and  would only
act as a global dielectric background as long as the excitation
energy remains small compared to their binding energy. 
In the present work, we obtain the energy loss 
spectrum for Ag using the same {\it ab initio} 
techniques but, this time, we focus on the plasmonic 
excitations of noble metals. In the case of Ag, 
the appearance of a plasma resonance at $\approx 3.8$
eV can be also qualitatively understood with help of a simple model,
which provides a new perspective on how a subthreshold plasmonic mode
can appear, and  what are the parameters controlling
the existence of this type of excitations. 

  The outline of the paper is as follows. In the following
section,  we discuss a model which provides a qualitative
explanation for some features in the energy loss spectrum up to
$\sim 10$ eV. The details of the {\it ab initio} computation of
the spectrum can be found in Sect.~\ref{abinitio}. A
description and discussion of the results is given in Sect.\ref{results}.
Finally, the main conclusions of the present paper are summarized in
Sect.~\ref{conclusions}. Atomic units ($\hbar = e^2 = m_e = 1$) are
used in all mathematical expressions.

\section{A simple model}\label{model}

 In order to estimate the plasma frequency in Ag  the classical
expression $\omega_p = \sqrt{4 \pi n}$ can be used. If we set
$n = n_c$, the density of conduction electrons, i.e., those
coming from the atomic $5s$ orbital, we find that $\omega^c_p = 8.98$ eV. 
However, if we furthermore include those coming from the atomic
$4d$ orbitals it turns out that $\omega_p = 32.6$ eV.
In both cases, we obtain values clear disagreement with the
experiment, which gives  a value of $3.8$ eV. Thus, as 
we have remarked in the previous section,  Ag does not seem to
behave as a ``simple'' metal whose plasmonic excitations can
approximately understood within the {\it jellium} model. 
Ehrenreich and Phillipp~\cite{EP62} considered this issue, 
concluding that it is also necessary to take
into account the  effect of electrons coming from the atomic $4d$ orbitals in a more
subtle way. In fact, it is necessary to account for the full band 
structure of Ag in order to explain what is observed in the
experiments. In what follows, we  shall study how the interband
transitions affect the collective motion of conduction electrons,
leading to the existence of a subthreshold plasmonic excitation.

 The usual way to find out the energy plasmonic excitations goes as follows.
Disregarding crystal local field effects,
plasmonic excitations correspond to zeroes in the
macroscopic dielectric function. In the $|{\bf q}| = 0$ limit
we have that:
\begin{equation}\label{meq1}
\epsilon(\Omega) = 0,
\end{equation}
for some complex $\Omega$. Using the Sellmeyer-Drude
expression~\cite{P65} for $\epsilon(\omega)$, and 
neglecting damping effects,
\begin{equation}\label{meq2}
\epsilon(\omega) = 1 - \frac{(\omega^{D}_p)^2}{\omega^2} +
 \delta\epsilon^{(ib)}(\omega),
\end{equation}
where last term corresponds to the interband contribution:
\begin{equation}\label{meq3}
\delta\epsilon^{(ib)}(\omega) = \sum_{\omega_n > 0}
    \frac{f_n}{\omega_n^2 - \omega^2}.
\end{equation}
The condition in Eq.~(\ref{meq1}) can be written as
\begin{equation}\label{meq4}
\Omega^2 - (\omega^{D}_p)^2 + \Omega^2 \delta\epsilon^{(ib)}(\Omega) = 0.
\end{equation}
The oscillator strength, $f_n$ in Eq.~(\ref{meq3}), is a measure of the effective
number of electrons that participate in the excitation of a given
mode `n', with energy $\omega_n$. The dominant contribution at
frequencies below $\omega^c_p$ comes from the excitation of conduction
electrons. The Drude plasma frequency~\cite{EP62} $\omega^{D}_p =
9.2$ eV, is slightly higher than $\omega^{c}_p$.
The ratio of the two plasma frequencies is usually
expressed in terms of the ``optical'' mass, $m_{opt}$ (recall that
 $m_e=1$ in atomic units):
\begin{equation}\label{meq5}
\left( \frac{\omega^{c}_p}{\omega^{D}_p} \right)^2  = m_{opt} = 0.95
\end{equation}
for Ag and $m_{opt} = 1.35$  for Cu.~\cite{EP62} In the RPA, $m_{opt}$
corresponds to the effective mass of electrons averaged
over the occupied part of the conduction band.

   Eq.~(\ref{meq4}) can be read in the following way. We assume that 
the term $\Omega^2 \delta\epsilon^{(ib)}(\Omega)$ plays the role of  a
``self-energy'' for the Drude plasmon. In simple metals, this
term is usually small and varies slowly with $\omega$.
This leads to a small shift of the Drude frequency. The plasma
frequency also acquires a small imaginary part, which accounts 
for the fact that the plasmon state is degenerate in energy 
with other configurations of the system
involving one or several electrons excited from one band to
another. This idea can be easily realized in a model in which the
plasmon is viewed as a boson coupled to an energy continuum. In an
independent electron picture, the continuum corresponds to the
electron-hole pair excitations.  A simple model 
Hamiltonian that can be set up
to describe this situation  is the following :
\begin{eqnarray}\label{meq6}
H  &=& \sum_{\bf q} \; \omega^{D}_p \:
       b^{\dagger}_{\bf q} b_{\bf q}   \nonumber\\
    &+& \sum_{k,{\bf q}} \; \omega_{k,{\bf q}}\:
        m^{\dagger}_{k,{\bf q}} m_{k,{\bf q}}\nonumber \\
    &+& \sum_{{\bf q},k} \;
               g_{k,{\bf q}} \, (b_{\bf q} + b^{\dagger}_{\bf -q})
     (m^{\dagger}_{k,{\bf q}} + m_{-k,{\bf -q}} ).
\end{eqnarray}
In this model, plasmons are presented as a field of oscillators carrying
momentum $\bf q$. Thus, $b^{\dagger}_{\bf q}$ and $b_{\bf q}$ are
plasmon creation and annihilation operators, respectively. To keep
the model as simple as possible, but retaining all the important
physical effects, the energy continuum is also represented
by a set of harmonic oscillators whose quanta are created and
annihilated by the operators $m^{\dagger}_{k,{\bf q}},\, m_{k,{\bf
q}}$. The label $k$ stands for the additional quantum numbers
(spin, relative momentum, band indices...) carried by a given
continuum mode.  The operator $m_{-k,{\bf -q}}$
($m^{\dagger}_{-k,{\bf -q}}$) anhilates (creates) a continuum mode with all
the momentum quantum numbers reversed. The assumed interaction 
between plasmons and the energy continuum conserves the momentum
$\bf q$. Terms connecting plasmons and modes with momentum
differing by a finite reciprocal lattice vector are therefore
neglected. This approximation amounts to disregarding crystal
local field corrections, which in general is a good approximation 
for metals. Indeed, in the frequency range
in which we are interested, namely $\omega \sim \omega^{D}_p$ , as
it can be seen from the {\it ab initio} calculations in Cu\cite{CRP99} and Ag (see
below), the local field effects at small ${\bf q}$ amount
to a few percent correction.

 The above model, Eq.~(\ref{meq6}), can be solved exactly.~\cite{G81,M90}
Indeed, it is related to the models introduced by Fano~\cite{F61} and
Anderson~\cite{A61} to study the coupling of a discrete state to
a continuum. More precisely, the present model can be thought of as an extension
of Gadzuk's work~\cite{G81} for a  localized vibrational mode 
in an electron liquid. Thus, the energies of the plasmonic
modes are given by the solutions of the following equation:
\begin{equation}\label{meq7}
\Omega^2 - (\omega^{D}_p)^2 + 2 \omega^{D}_p M({\bf q},\Omega) = 0,
\end{equation}
where $M({\bf q}, \omega)$ is the plasmon self-energy. It can be
expressed in terms of its imaginary part:
\begin{eqnarray}\label{meq8}
{\rm Im}\; M({\bf q},\omega) = \pi
   \sum_{k} | g_{k,{\bf q}}|^2 \: [&&
      \delta (\omega - \omega_{k,{\bf q}}) \nonumber \\
    &&  - \delta(\omega + \omega_{k,{\bf q}}) ],
\end{eqnarray}
by means of the expression:
\begin{equation}\label{meq9}
M({\bf q},\omega) = \int \; \frac{{\rm d}\omega'}{\pi}\,
  \frac{{\rm Im}\; M({\bf q},\omega')}{\omega' - \omega - i\eta},
\end{equation}
with $\eta \to 0^+$.

Working by analogy with Eq.~(\ref{meq4}), leads us to tentatively
identify
\begin{equation}\label{meq10}
M(|{\bf q}| \to 0,\omega ) =
  \frac{\omega^2}{2 \omega^{D}_p} \delta \epsilon^{(ib)}(\omega).
\end{equation}
If the RPA expression of $\delta\epsilon^{(ib)}(\omega)$ is used,
we can express it in terms of single-electron
orbitals, $|n\: {\bf k}  \rangle$, and energies,
$\varepsilon_{n {\bf k}}$:
\begin{eqnarray}\label{meq11}
 {\rm Im}&& M(|{\bf q}|\to 0, \omega) =
\frac{2 \pi^2}{3 V \omega^D_p}  \sum_{{\bf k} \in BZ} \sum_{m \neq n}
\, f(\varepsilon_{n {\bf k }}) \nonumber \\
&& \times  \left[ 1 - f(\varepsilon_{m {\bf k }} ) \right]
 | \langle m \: {\bf k} | {\bf p} | n\: {\bf k}  \rangle |^2
 \delta(\omega - \varepsilon_{m {\bf k }} + \varepsilon_{n {\bf k }} ),
\end{eqnarray}
where $V$ is the volume of the system and ${\bf p}$ the momentum
operator. We have assumed cubic symmetry and a local pseudopotential; 
if the pseudopotential is not local, the
matrix element of $\bf p$ must be replaced by a more complicated
expression. For $\omega = \omega^{D}_p$ this equation matches the
Golden Rule result for the plasmon line width in the $|{\bf q}|=
0$ limit,~\cite{BB94} which confirms the identification made
above in Eq.~(\ref{meq10}). For arbitrary $\omega$, however, it is
better to look at ${\rm Im}\; M({\bf q},\omega)$ 
as a weighted density of the continuum modes, cf. Eq.~(\ref{meq8}). 
Thus, it follows from Eqs.~(\ref{meq7}) and (\ref{meq9}) 
that if $\omega^{D}_p$ lies above (below) the region where 
${\rm Im}\; M({\bf q},\omega)/\omega^D_p$ is large 
(provided  ${\rm Im}\, M({\bf q},\omega^D_p)/\omega^D_p$ be small), the 
plasma frequency will be blue (red) shifted. This
result is in agreement with the above mentioned work of Wilson.~\cite{W60}
When ${\rm Im}\, M({\bf q},\omega^D_p)/\omega^D_p$  is
large, we will not observe a well-defined plasma resonance
in the spectrum.  This seems to be the case of noble metals due to
the presence of a manifold of occupied $d$-like
bands, which gives rise to a broad band of modes extending
from the interband threshold energy  $\omega_T \approx 3.9$  eV 
(for Ag and $2.1$ eV for Cu), to well above the Drude
plasma frequency ($9.2$ eV for Ag and $9.3$ eV for Cu). 
If we assumed that the onset of interband
transitions in Ag took place so sharply as to produce a
discontinuity in the weighted density of continuum modes:
\begin{equation}\label{meq12}
{\rm Im}\, M(|{\bf q}|\to 0, \omega) \sim \theta(\omega_T - \omega)
\end{equation}
for $\omega \sim \omega_T$, then ${\rm Re}\, M({\bf q}|\to 0,
\omega)$ would develop a logarithmic behavior $\sim -
\log(\omega - \omega_T)$ for $\omega < \omega_T$. This means that
Eq.~(\ref{meq7}) would have a dampingless solution just below the
threshold, $\omega_T$. This polaronic-like solution~\cite{ES63} 
corresponds to an hybrid plasmonic mode, in which the conduction
electrons oscillate coherently without exciting the 
electrons in the $d$-like bands. These are just polarized by the 
electric field set up by plasmonic mode, 
so that the plasma frequency is lowered.  
This point of view agrees well with more phenomenological
approaches,~\cite{L93,SR97} which simply assume that the red shift in the plasma frequency
can be accounted for by an effective dielectric function $\epsilon_d$, for
the electrons in the $d$-like bands so that $\omega^{*}_p = 
\omega^D_p/\sqrt{\epsilon_d} \approx 3.8$ eV.

 Additionally, there is another complex solution of Eq.~(\ref{meq7}), 
for which the conduction electrons  move as an overdamped oscillator, 
rapidly decaying into an electron hole-pair. For $|{\bf q}| \to 0$, this 
excitation corresponds to an optical interband transition. 
Two types of optical  transitions  may occur above the 
threshold.~\cite{MP67} Two examples have been indicated by arrows in Fig.~\ref{fig1}, which
represents the calculated first-principles LDA band structure of Ag.
Type A corresponds to the excitation of an electron from one
flat $d$-like band to an unoccupied state in the conduction band or above. The other
type (B), whose threshold occurs at approximately the same energy in the three
noble metals, corresponds to the excitation from an occupied
state in the conduction band (near $L$ point in the BZ) to the flat
part of an unoccupied band or above. This type of transitions 
seems to be responsible for the
damping of the plasmon when the $4d$ electrons are
included in the pseudocore, i.e., they are considered to be frozen.
In Sect.\ref{results}, we shall explain this issue more in detail. 
These two types of transitions are important because  
they often involve flat bands, which may lead to van Hove singularities in the
joint density of states.

 In the case of Ag, the onsets for the transitions of type A and B
nearly overlap in energy. This produces a sharp interband onset and
gives rise to the pronounced peak  that ${\rm Im}\,
\epsilon(\omega)$ exhibits around $4$ eV. 
From Eq.~(\ref{meq10}), it follows then that ${\rm Im}\, M(|{\bf q}|\to 0, \omega)$ will also 
have a maximum at the same energy. The onset of the interband transitions at
$\omega_T$ is not as sharp as suggested by Eq.~(\ref{meq12}).
However, it still leads to a zero of $\Omega^2 - (\omega^D_p)^2  + 2 \omega^D_p {\rm
Re}\,M(|{\bf q}|\to 0, \Omega)$ for a real $\Omega$, which occurs
below the threshold, where the density of continuum modes is very small.
This point is illustrated by Fig.~\ref{fig2}, which displays 
the graphical  solution of Eq.~(\ref{meq7}). 
This provides an explanation for the
existence of a subthreshold plasma resonance in Ag. However,  it remains to
be explained why a similar phenomenon is not observed in the
spectra of Cu or Au. The difference stems from the specific details of the
band structure and, in particular, from the value of the threshold energy. In
Fig.~\ref{fig3}, we have plotted the graphical solution of Eq.~(\ref{meq7})in 
the case of Cu. The threshold for the excitation of $d$-like bands 
occurs this time at a lower energy ($2.1$ eV) because the onset for the
interband transitions of type A and B does not overlap. 
Comparison of Figs.~\ref{fig2} and \ref{fig3} shows the
important role played by the value of the threshold energy, 
cf. Eq.~(\ref{meq10}), in enhancing the maximum of ${\rm Im}\, M(|{\bf q}|\to 0, \omega)$,
and thus in ${\rm Re}\, M(|{\bf q}|\to 0, \omega)$.
Therefore, for Cu no  zero of Eq.~(\ref{meq7}) close to the real axis exists
below the threshold, where this plasma mode would have a small
density of continuum modes to decay. A similar conclusion can be drawn 
for the case of Au. 

  To sum up, band structure effects  
modify the original Drude plasmon, which 
turns out to be no longer a well defined plasmonic excitation. Thus, 
it shows up in the Ag spectrum as broad peak with a maximum around $8$ eV.
Moreover, a combination of a high value for $\omega_T$
along with the sudden onset of interband transitions  
produces a narrow plasma resonance in Ag. In this respect, our conclusions agree with those of Ehrenreich
and Phillipp.~\cite{EP62} Furthermore, the line width of this ``delicate'' feature in
the energy loss spectrum also depends on the sharpness of the interband onset. 
The sharper onset the narrower the resonance, which in any case has very little
oscillator strength.  This makes this plasmonic excitation very
sensitive to changes in the band structure, 
impurities or defects. This is consistent
with the experimental data from alloying experiments,~\cite{W72,ML68,IHW70}
which indicate that the plasmon is strongly damped by increasing the concentration
of the other component in the alloy. Finally, it is worth pointing out 
that the existence of this resonance also depends on the value of the Drude 
plasma frequency. If the value of $\omega^D_p$ is increased,
for example by increasing the electron density in the system, the
resonance could disappear from the spectrum.
This prediction is confirmed by our {\it ab initio} results (shown below) 
for the energy loss spectrum of Ag under externally applied
hydrostatic pressure.

\section{Details of the  {\it ab initio} calculations}\label{abinitio}

 By measuring the energy lost by electrons or
X-ray photons in their interaction with matter, one can access the
dynamical structure factor, $S({\bf q},\omega)$ of a system. The
fluctuation-dissipation theorem~\cite{PN66} allows us to relate
$S({\bf q},\omega)$ to the density correlation function:
\begin{equation}\label{aeq1}
\chi({\bf r}, {\bf r'}, t) = -i \theta(t) \;
  \langle \left[ \delta n_H({\bf r}, t), \delta n_H({\bf r}',0)
        \right] \rangle,
\end{equation}
where  $\delta n_H({\bf r}, t) = n_H({\bf r}, t) - \langle n_H({\bf
r}, t) \rangle$, and suffix $H$  means that  operators are in
the Heisenberg picture~\cite{M90}; the brackets mean that the average
is taken over the ground state of the system. Now, time-dependent 
density functional theory~\cite{tddft} (TDDFT) provides
us with a method to compute this correlation 
function by solving the following integral equation:
\begin{eqnarray}\label{aeq2}
&& \chi_{{\bf G}{\bf G'}}({\bf q},\omega)
   = \chi^0_{{\bf G}{\bf G'}}({\bf q},\omega)
    + \sum_{{\bf G}_1,{\bf G}_2} \,
       \chi^0_{{\bf G}{\bf G}_1}({\bf q},\omega) \nonumber \\
&& \times \left[ v_{{\bf G}_1{\bf G}_2}({\bf q})
 +  K^{xc}_{{\bf G}_1{\bf G}_2}({\bf q},\omega) \right]
     \chi_{{\bf G}_2{\bf G'}}({\bf q},\omega).
\end{eqnarray}
Here, we have exploited crystal symmetry and time translation invariance  
by introducing:
\begin{eqnarray}\label{aeq3}
\chi_{{\bf G}{\bf G'}}({\bf q},\omega) = \frac{1}{V} 
 \int {\rm d}t \; &&
  {\rm d}^3{\bf r} \; {\rm d}^3{\bf r'} \,
  e^{i \omega t}
  e^{-i({\bf q} + {\bf G})\cdot {\bf r}} \nonumber \\
 && \times e^{i({\bf q} + {\bf G'})\cdot {\bf r'}}\;
  \chi({\bf r}, {\bf r'},t),
\end{eqnarray}
where $\bf q$ belongs to the BZ, and
${\bf G},{\bf G'}$ to the reciprocal lattice.
Similar definitions hold for the other functions that appear in
Eq.~(\ref{aeq2}). For example,
\begin{equation}\label{aeq4}
v_{{\bf G}{\bf G'}}({\bf q}) =
  \frac{4\pi}{|{\bf q} + {\bf G}|^2} \delta_{{\bf G},{\bf G'}}
\end{equation}
corresponds to the Fourier components of the Coulomb interaction between metal
electrons. The function
\begin{eqnarray}\label{aeq5}
\chi^0_{{\bf G}{\bf G'}}({\bf q},\omega) = \frac{1}{V}\:
  \sum_{{\bf k} \in {\rm BZ}} && \sum_{n,m}\, 
  \frac{ f(\varepsilon_{n{\bf k}}) - f(\varepsilon_{m{\bf k}+{\bf q}}) }
    {\omega + \varepsilon_{n{\bf k}} - \varepsilon_{m{\bf k}+{\bf q}}
 + i \eta} \nonumber  \\
&& \times \langle n\: {\bf k} | e^{-i({\bf q} + {\bf G})\cdot {\bf r}}
 | m\: {\bf k} + {\bf q} \rangle \nonumber  \\
&& \times
\langle m\: {\bf k} + {\bf q} | e^{i({\bf q} + {\bf G'})\cdot {\bf r}}
  | n\: {\bf k}  \rangle
\end{eqnarray}
with $\eta \to 0^+$, is the density correlation function
for a system of independent particles, with single-particle energies
$\varepsilon_{n{\bf k}}$ and orbitals  $|n\: {\bf k} \rangle$.
These are eigenvalues and eigenstates of a single-particle
Hamiltonian. The potential energy in this Hamiltonian is given by
the Kohn-Sham~\cite{KS65} potential. In our case, the Kohn-Sham potential has been
calculated within LDA~\cite{LDA,KS65}. 

  We have used a plane wave basis to expand the Bloch eigenstates.
For this procedure to be computationally efficient, it is
necessary to replace the lattice potential by a superposition of
ionic pseudopotentials having the same scattering properties. 
However, as electrons in the outer atomic $4d$ shell play 
an important role in the response of
noble metals, they are treated as a part of the valence electrons. The  
remaining core electrons are replaced by a frozen pseudocore
using a norm-conserving pseudopotential scheme.~\cite{TM91}
Therefore, electrons included in the pseudocore 
will not respond to an external perturbation. 
This is not an important drawback as long as the
excitation energy remains below the typical excitation
energies of core electrons $\sim 100$ eV.

 Including $4d$ electrons in the valence produces a pseudopotential
that requires a relatively high number of plane waves to get good
convergence. Here, we  have used a scalar relativistic
pseudopotential generated according to the scheme
of Troullier and Martins.~\cite{TM91} It requires $\approx 1,400$
plane waves per Bloch state up to a cutoff energy of $\approx 1.1$ keV 
in order to obtain a well converged ground state energy (less than $0.01$ eV), 
and single-electron energies and orbitals 
up to $\approx 100$ eV. The lattice constant has been set to
the experimental value of the FCC crystal structure of Ag,
namely $a = 4.09\, {\rm \AA}$. We have also solved the Kohn-Sham equations 
with $a = 3.70\, {\rm \AA}$, which corresponds to an 
externally applied hydrostatic pressure of $63$ GPa.

 The RPA is equivalent to solving Eq.~(\ref{aeq5})
with  $K^{xc}=0$. Therefore,
all the exchange correlation effects beyond the RPA
are contained in the kernel $K^{xc}_{{\bf G}{\bf G'}}({\bf q},\omega)$.
Consistently with the use of LDA eigenstates and eigenvalues in
the calculation of $\chi^0$, this
function is approximated by the following expression:
\begin{equation}\label{aeq6}
K^{xc}_{{\bf G}{\bf G'}} =  \frac{1}{V}
 \int {\rm d}^3{\bf r} \; e^{- i ({\bf G} - {\bf G'}) \cdot {\bf r} } \;
f_{xc}({\bf r}),
\end{equation}
where
\begin{equation}\label{aeq6b}
f_{xc}({\bf r}) = \left[ \frac{dV_{xc}(n)}{dn}\right]_{n_{gs}({\bf r})}.
\end{equation}
Here $V_{xc}$ is the exchange correlation part of the
Kohn-Sham  LDA potential, and $n_{gs}({\bf r})$ is the ground state
density. This approximation  is usually called {\it
adiabatic} local density approximation (ALDA). However, it is
observed that for metals in the long wavelength limit, ALDA and
RPA yield very similar results. This has been found in simple
metals,~\cite{KE99,QE93} and  also in Cu.~\cite{CRP99} 
Thus, when computing the
spectrum, we have used the RPA.

 Instead of solving Eq.~(\ref{aeq5}), we have evaluated the
RPA dielectric matrix:
\begin{equation}\label{aeq7}
\epsilon_{{\bf G}{\bf G'}}({\bf q},\omega) =
\delta_{{\bf G'},{\bf G'}} - v_{{\bf G}{\bf G'}}({\bf q})
  \chi^0_{{\bf G}{\bf G'}}({\bf q},\omega).
\end{equation}
In the RPA, the dynamic structure factor $S({\bf q,\omega})$ is  proportional
to the loss function, ${\rm Im}\,\left[-\epsilon^{-1}_{\bf 0
0}({\bf q},\omega)\right]$, which is the ${\bf G} = {\bf G'} = {\bf 0}$
element of the inverse dielectric matrix.  To calculate the dielectric matrix
we have used a cutoff of $41$ eV in ${\bf G},{\bf G}'$, which amounts to
using $59\times59$ matrices.  

 To  evaluate  $\chi^0$ using Eq.~(\ref{aeq5}), 
we have cut off the sum over $n$ and $m$ using
bands up to an energy of $\approx 100$ eV. The integration over 
the first BZ has been performed using a $20\times 20\times 20$ Monkhorst-Pack
mesh~\cite{MP76} in the irreducible BZ, which allows to deal 
effectively with the topology of the Fermi surface of Ag. Indeed, 
this is a very dense mesh and the
calculation of $\chi^0$ becomes certainly time and memory consuming. However, if
one is to resolve the very narrow plasma resonance that Ag
presents near the interband excitation threshold, it is then
necessary to work with such a fine mesh. 

In order to further reduce the numerical damping 
$\eta$ we have
also computed $\chi^0$ on the imaginary frequency axis by
making the replacement $\omega + i \eta \to i \nu$ in Eq.~(\ref{aeq5}). 
Analytic continuation to the real axis has been carried out by using a Pad\'e
approximant.~\cite{LC96} This procedure allows us to use a
numerical damping as small as $0.001$ eV. This is to be compared
with the $0.2$ eV value used in the real frequency calculations
also presented in this work. However, we  shall show the 
results of boht types of calculations together, to
indicate that one must be very careful with the analytic
continuation procedure using Pad\'e approximants. Thus, 
the results obtained using this method tend to be 
smoother than what is obtained with a calculation with real
frequencies and larger values of $\eta$. Thus, it usually happens
that some features of the spectrum, as obtained in a real frequency calculation,
are completely missing from the
analytically continued result (see following section). In fact, 
although we have used Lentz algorithm~\cite{PTVF92}
to evaluate the corresponding continued fraction, we have been unable to ensure that the procedure
converges by increasing the number of points over the same interval of $\nu$. 
When fitting the Pad\'e approximant~\cite{points} to the calculated values of 
$\epsilon_{{\bf G},{\bf G}}({\bf q}, i\nu)$, we have  
found that the results are unstable with respect to increasing
the number of points. Thus, we have obtained differences of
the order of $10$ \% when the number of points over the same
interval was simply doubled. Finally, we have
decided to use the modulus of the remainder 
of the truncated continued fraction
to estimate  the error, and always compare the outcome with a
real frequency calculation prior to consider it as reliable.

\section{Results}\label{results}

 In Fig.~\ref{fig1} we have plotted the band structure of 
Ag as calculated within the LDA. Two examples of the main types 
of interband transitions (A and B) which couple 
to  plasmons at energies $\omega \sim \omega^D_p$ are indicated by arrows.
In the LDA, we obtain a threshold
energy for  type A transitions $\approx 3$ eV, whereas experimentally
it is found $4$ eV.  Since  LDA eigenvalues enter the expression for $\chi^0$,
this turns out to be the reason for underestimating  the
threshold frequency, $\omega_T$, as shown in
Figs.~\ref{fig4} and \ref{fig5}. An underestimate
of this kind was already observed for Cu, where $\omega_T \approx
1.5$ eV in contrast with the experimental value of $2.1$ eV.
The source of such an underestimate may be
either a failure of the LDA or an indication that a frequency dependent and perhaps 
nonlocal $K^{xc}$ may be required to fully account for the
experimental value of $\omega_T$. In our opinion, however, it
is related to a failure of the LDA, which may not
be taking into account the strong correlations occurring amongst
the electrons in the flat $d$-like bands of Ag. Therefore, a more detailed
description of correlation for the $d$-like bands seems to be
required.\cite{Jorge}

Fig.~\ref{fig4} presents our results for the energy loss function
for  ${\bf q} = (0.05\, 0\, 0) (2 \pi / a)$ as obtained from the
LDA band structure. Since our interest is focused on the plasmonic excitations,
which occur for small ${\bf q}$, we need not consider exactly the 
$|{\bf q}| = 0$ limit. Thus, for the small $\bf q$ considered, 
the theory developed in Sect.~\ref{model} still applies. 
The continuous line  in this figure corresponds to 
${\rm Im}\:\left[-\epsilon^{-1}_{\bf 0 0}({\bf q},\omega)\right]$, whereas the
dashed line corresponds to ${\rm Im}\, \left[-1/\epsilon_{\bf 0 0}({\bf
q},\omega)\right]$, i.e. with and without including crystal
local field effects respectively. As we have already said
in Sect.~\ref{model}, crystal local field effects are at most
a correction $\sim 10\,\%$ in the considered frequency range. Dots are the experimental
data~\cite{P85} obtained by means of optical measurements (i.e., for
$|{\bf q}|  \to 0$), and  should be compared with the dashed line. 
As a consequence of  the LDA underestimate of $\omega_T$, 
the plasmon peak appears at a lower energy
$\approx 3.4$ eV than observed experimentally~\cite{EP62} 
$\approx 3.8$ eV. This peak also looks more damped either
because the high value of the numerical damping 
employed in the calculation and  because 
of the LDA underestimate of $\omega_T$.

 From Figs.~\ref{fig4} and \ref{fig5} we can conclude that: 
(i) The Drude plasmon becomes so broadened  by
its coupling to interband transitions that we can no longer speak of
a well defined excitation. The high energy tail of this broad
feature around $8$ eV is more strongly affected by crystal local 
field effects. This seems to  indicate that those decay processes 
involve the excitation of  an electron in a more localized state,
probably lying deeper in the $d$-like band. The important role played
by the $d$-like bands in destroying the Drude plasmon can be seen in
Fig.~\ref{fig6}, where the dashed line corresponds to a calculation
of the energy loss function putting the ten $4d$ electrons of atomic 
Ag in the frozen pseudocore. In this case, the Landau damping of the
Drude plasmon is only due to transitions of type B, which have a much
smaller spectral weight than those of type A. Furthermore, the 
plasma frequency is pushed upwards by this continuum of modes, which
mainly lies below $\omega^D_p$. However, when electrons from the $d$-like
bands are considered as active, the effect goes in the opposite direction
thus shifting the maximum of the Drude peak  from $\omega^D_p$ eV to $\approx 7.5$ eV; 
(ii) In agreement with the experiments, 
a double hump structure in the region between 15 to 30 eV is found. 
This is a band structure effect, which turns out to be a general feature 
in the noble metal systems.~\cite{CRP99} 
As it can be seen in Fig.~\ref{fig5}, although the
calculation using a Pad\'e
approximant (dot-dashed line) gives a reliable description  of 
the low frequency part, it fails to reproduce the double hump structure. 
This seems to be  a rather general feature of analytic continuation 
by Pad\'e approximants.  Whenever the function to be continued 
presents two poles close to each other and having a large imaginary part,
the values for the function on the real axis correspond  to an
average of the two poles, and a single broad peak shows up in
the outcome of the analytical continuation.  (iii) Altogether, 
the {\it ab initio} LDA calculations presented here give a
good description of the high energy features (i.e. above $\omega^D_p$)
of the energy loss spectrum of Ag. These include the 
Landau damping of the Drude plasmon and
the double hump structure mentioned above. This could be expected
on the basis that the spectral weight 
in the high energy region of the spectrum comes, 
to a large extent, from the excitation
of electrons to bands well above the Fermi level, which are   
expected to be more accurately reproduced by the LDA,
as it is an approximation suitable for an homogeneous
electron system. However, the LDA underestimate
of the threshold energy is responsible for 
failing to reproduce the low energy structure.
This issue is further investigated below.

 In Fig.~\ref{fig6}, we show the results for the real
and imaginary parts of $\epsilon_{\bf 0 0}({\bf q}, \omega)$
for ${\bf q} = (0.05 \, 0 \, 0) (2 \pi /a)$. The same remarks
made above apply in this case as well. The experimental
results, full circles for the imaginary part and  open squares for the real part
are from Ref.~\onlinecite{P85}. In this figure we can see more 
easily that the onset for the excitation of interband transitions
occurs, not only at a lower energy, but that it is less sharp as well.
As it has been remarked in the previous section, this leads to a 
more damped plasmonic mode below the threshold.

  In order to investigate the influence of the interband threshold
energy on the existence of a subthreshold plasma resonance,  we
have performed a series of numerical ``experiments''. The
motivation of these experiments is to understand why Ag presents a
narrow plasma resonance in its energy loss spectrum, while
Cu (and Au) does not, despite the similarity of their band structures. 
To modify the threshold energy $\omega_T$ without opening unphysical
gaps in the band structure, we have scaled by a factor $\alpha$
the separation between the Fermi energy ($\varepsilon_F$) 
and the LDA eigenvalues of the occupied states:
\begin{equation}\label{aeq8}
\varepsilon'_{n{\bf k}} = \varepsilon_F +
\alpha (\varepsilon_{n{\bf k}} - \varepsilon_F).
\end{equation}
Thus, using this new set of eigenvalues, we have calculated $\chi^0$
and hence obtained the energy loss function. This rather {\it ad hoc} 
procedure does not modify the wave functions accordingly. Furthermore, it
introduces a discontinuity of $\nabla_{\bf k} \varepsilon'_{n{\bf k}}$ at the 
Fermi surface, so that it might lead unphysical results. However, 
as shown in Fig.~\ref{fig7},
this seems not to be the case, at least for $\omega \sim \omega_T$ and 
$\bf q$ finite. As predicted in Sect.~\ref{model}, the subthreshold 
plasmon peak is enhanced and shifts up in energy 
as $\omega_T$ increases. In this figure, continuous lines 
correspond to calculations for real frequencies with 
$\eta = 0.2$ eV, while dashed lines
correspond to the analytic continuation using a Pad\'e approximant 
from the results obtained on the imaginary frequency axis. 
Both calculations show the same tendency. On the other hand, when
the threshold energy is further decreased, from its LDA value $\approx 3$ eV,
to $\omega_T \approx 1.5$ eV,  the plasma resonance disappears 
and the spectrum becomes more ``Cu-like''.

  Finally, in Fig.~\ref{fig8} we show the  energy loss spectrum of Ag
under an external pressure of 63 GPa. As in previous cases, the
continuous line corresponds to a calculation using real frequencies
while the dashed line is the result of an analytical continuation using
a Pad\'e approximant. Notice that in this case the
lattice constant is $a_{press} = 3.7\, {\rm \AA}$ so that now $q = (0.05\,  0 \, 0)
(2 \pi / a_{press})$, corresponds to a slightly larger momentum
($|{\bf q}_{press}| = 8.5 \times 10^{-2}\, {\rm \AA}^{-1}$ vs 
$|{\bf q}| = 7.7 \times 10^{-2}\,{\rm \AA}^{-1}$). 

Decreasing the lattice constant changes the width of the bands so that they
disperse more rapidly with $\bf k$ (i.e., the band width becomes larger).
Furthermore, electronic density also increases. This shifts the
maximum previously at $\omega^{max} \approx 7.5$ eV to 
$\omega^{max}_{press}\approx 8.8$ eV. Since this maximum
is related to the Drude plasmon, whose {\it bare} frequency scales with
the lattice parameter as $a^{-3/2}$, we could understand the shift 
(neglecting the change in the optical mass) as due to the 
increase in the electr\'onic density. Indeed, this argument
works reasonablely well as $(a_{press}/a)^{-3/2} = 1.16$, and
$\omega^{max}_{press}/\omega^{max} \approx 1.17$~. Moreover, as the 
bands become wider, specially those corresponding to excited states,
more oscillator strength is transferred to higher energies.
This decreases the density of modes at the Drude frequency, 
and thus the corresponding peak is now less broad. What is more, in this spectrum
the subthreshold plasmon is not present any more. Again,
this is due to the increased electronic density, which  yields
a higher value for the Drude plasma frequency so that Eq.~(\ref{meq7})
has no longer a solution close to the real axis. The behavior 
under pressure predicted here remains to be confirmed by experiments.

\section{Conclusions}\label{conclusions}
 
  Here, we summarize the main conclusions of the present work.
We have shown that a first principles calculation of the energy loss
spectrum is able to reproduce, within the RPA, the high energy structure
in the spectrum of Ag. However, since the LDA underestimates the
interband energy threshold, the agreement is not as good as for the high
energy region around $4$ eV. For example, the plasma resonance observed
experimentally at $3.8$ eV appears as  a more damped peak at $3.4$ eV.
In our opinion, this underestimate may be related to a failure of the LDA to
account for  electron correlation effects that take place in the flat d-like
bands of Ag. In fact, an {\it ad hoc} scaling of the occupied bands
so that the threshold energy is brought closer to the 
experimental value gives rise to a better defined
subthreshold plasmonic excitation. We  have also shown 
the important role played by the $d$ symmetry
bands in determining the existence of the plasma resonance at $3.8$ eV, just
below the interband excitation threshold. By using a simple model, it is shown
that the appearance of this excitaci\'on has to do with the sharpness of
the interband onset and the relative values of the threshold energy and the Drude
plasma frequency. Since it contains very little oscillator strength, it is
very sensitive to impurities, defects, and changes in the band
structure. Thus, we have demonstrated by performing numerical experiments that its presence
depends on the value of the threshold energy. 
Finally, we have also computed the energy loss spectrum
of Ag under pressure, and found that its features can
be aslo understood in terms of the simple model presented here. We remark that
the techniques used and analysis carried out in this work are not specific
of Ag, and could be readily extended to other metallic compounds with
fully occupied $d$ symmetry bands close to the Fermi level. 

\acknowledgements

 We thank N. Lorente,  J.~F. Dobson, and L. Hedin  for fruitful
discussions, and I. Campillo for many useful suggestions. 
This work was supported by the Basque {\it Unibertsitate
and Hezkuntza Saila}, JCyL (VA28/99), and {\it Iberdrola, S.A}. 
M.A.C. acknowledges a scholarship granted from the Basque Government, and
J.S.D. a grant from The University of Basque Country/{\it Euskal Herriko Unibertsitatea}.
M.A.C. and J.S.D. wish to acknowledge the kind hospitality of the {\it Departamento de
F\'{\i}sica Te\'orica} in Valladolid.


\begin{figure}
\caption[]{ \label{fig1}
Calculated  LDA band structure of Ag. Arrows indicate the two examples (A and B)  
of the  types  of interband transitions that couple to the 
plasmonic excitations in the long wavelength limit, for energies above the interband
excitation threshold. The crystal structure is FCC with lattice parameter equal to
$a=4.09\, {\rm \AA}$. The zero energy corresponds to the Fermi level.
}
\end{figure}
\begin{figure}
\caption[]{ \label{fig2}
Graphical solution of Eq.~(\ref{meq7}) for the subthreshold plasmonic modes in Ag. 
Data for $\delta \epsilon^{(ib)}(\omega)$
(i.e., for ${\rm Re}\; M(|{\bf q}| \to 0,\omega)$) are from Ref.~\onlinecite{EP62}. The 
continuous line cuts the dashed line at the point indicated by the black dot. This point
lies just below the threshold, where the density of continuum modes is small. As
a consequence, a narrow  plasma resonance appears in the spectrum.
}
\end{figure}
\begin{figure}
\caption[]{ \label{fig3}
Graphical solution of Eq.~(\ref{meq7}) for the subthreshold plasmonic modes in Cu. 
Data for $\delta \epsilon^{(ib)}(\omega)$
(i.e., for ${\rm Re}\; M(|{\bf q}| \to 0,\omega)$) are from Ref.~\onlinecite{EP62}. The 
continuous line does not cut the dashed line, indicating that no subthreshold plasmonic
modes exist in Cu. A similar situation is expected to hold for Au.
}
\end{figure}
\begin{figure}
\caption[]{ \label{fig4}
Energy loss spectrum of Ag for $q=(0.05\,0\,0)(2\pi/a)$,
$|{\bf q}| = 7.7\times 10^{-2}\,{\rm \AA}^{-1}$. The continuous line is imaginary part of
${\bf G} = {\bf G'} = {\bf 0}$ element of the inverse dielectric matrix,
namely the energy loss function
${\rm Im}\, \left[ - \epsilon^{-1}_{\bf 0 0}({\bf q},\omega)\right]$, 
whereas the dashed line corresponds to (the imaginary part of) 
one over the  ${\bf G} = {\bf G'} = {\bf 0}$
element of the dielectric matrix
${\rm Im}\, \left[-1/\epsilon_{\bf 0 0}({\bf q},\omega) \right]$. Differences
between them are due to crystal local field effects. Dots are the experimental
data from Ref.~\onlinecite{P85}. As they were obtained from optical data and do
not include crystal local field corrections, they
must be compared to the dashed line. 
}
\end{figure}
\begin{figure}
\caption[]{\label{fig5}
Energy loss spectrum of Ag for  $q=(0.05\,0\,0)(2\pi/a)$,
$|{\bf q}| = 7.7 \times 10^{-2}\,{\rm \AA}^{-1}$. The continuous line
is a calculation for real frequencies with $\eta = 0.2$ eV, cf. Eq.~(\ref{aeq5}).
The dot-dashed line, however, is obtained by analytically  continuing to the real axis
an imaginary frequency calculation, using a Pad\'e approximant. As it can be seen, the
results obtained by analytical continuation cannot reproduced all the features in the 
spectrum obtained using a real $\omega$ (but higher numerical damping $\eta$). The 
dashed line is a calculation of the loss function including the $4d$ electrons of Ag in
the frozen pseudocore that replaces the true ionic core. Then, these electrons cannot be excited,
thus showing the important role that they actually play in modifying the 
spectrum. Notice that the intensity  of this peak is five times what is represented here.}
\end{figure}
\begin{figure}
\caption[]{ \label{fig6}
Dielectric function $\epsilon_{\bf 0 0}({\bf q},\omega)$
 of Ag for $q=(0.05\,0\,0)(2\pi/a)$,$|{\bf q}| = 7.7 \times 10^{-2}\,{\rm \AA}^{-1}$.
The continuous line corresponds to the imaginary part as obtained by analytically continuing
an imaginary frequency calculation, using a Pad\'e approximant. The dashed line is the real
part obtained by the same method. We checked that the results are consistent with a real
frequency calculation (not showed here for clarity). Full circles (imaginary part) and
open squares (real part) correspond to experimental data form Ref.~\onlinecite{P85}.
}
\end{figure}
\begin{figure}
\caption[]{ \label{fig7}
Results of the scaling procedure described in Sect.~\ref{results}. The continuous line
are the results as calculated for real $\omega$ with $\eta = 0.2$ eV, while the 
dashed lines are obtained by analytical continuation using  Pad\'e approximants.
When the occupied bands are scaled so that the interband threshold energy increases,
the subthreshold plasmon appears more defined. However, when threshold energy 
is decreased the plasmon disappears from the energy loss spectrum.
}
\end{figure}
\begin{figure}
\caption[]{ \label{fig8}
Energy loss spectrum of Ag for  $q=(0.05\,0\,0)(2\pi/a_{press})$ 
($|{\bf q}| = 8.5\times10^{-2}\,{\rm \AA}^{-1}$) with $a_{press}=3.7\, {\rm \AA}$,
which corresponds to an externally applied hydrostatic pressure of $63$ GPa. 
The continuous line is a calculation for real frequencies  with $\eta = 0.2$ eV, 
cf. Eq.~(\ref{aeq5}). The dashed line, however, is obtained 
by analytically  continuing to the real axis
an imaginary frequency calculation, using a Pad\'e approximant.
}
\end{figure}


\begin{thebibliography}{30}
\bibitem{points}
We have used up to 1600 points over up to 110 eV in $\nu$.
\end{thebibliography}
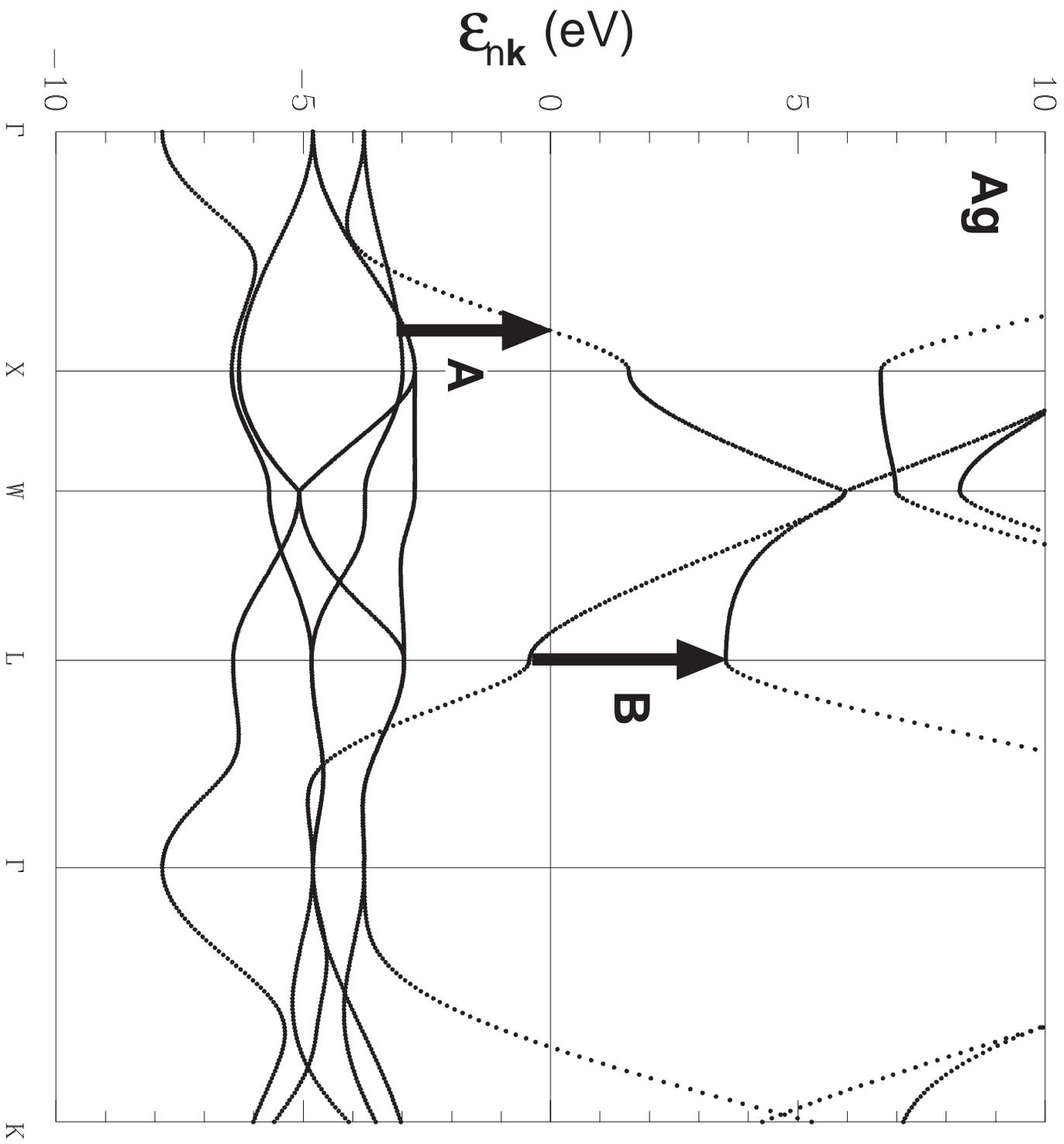
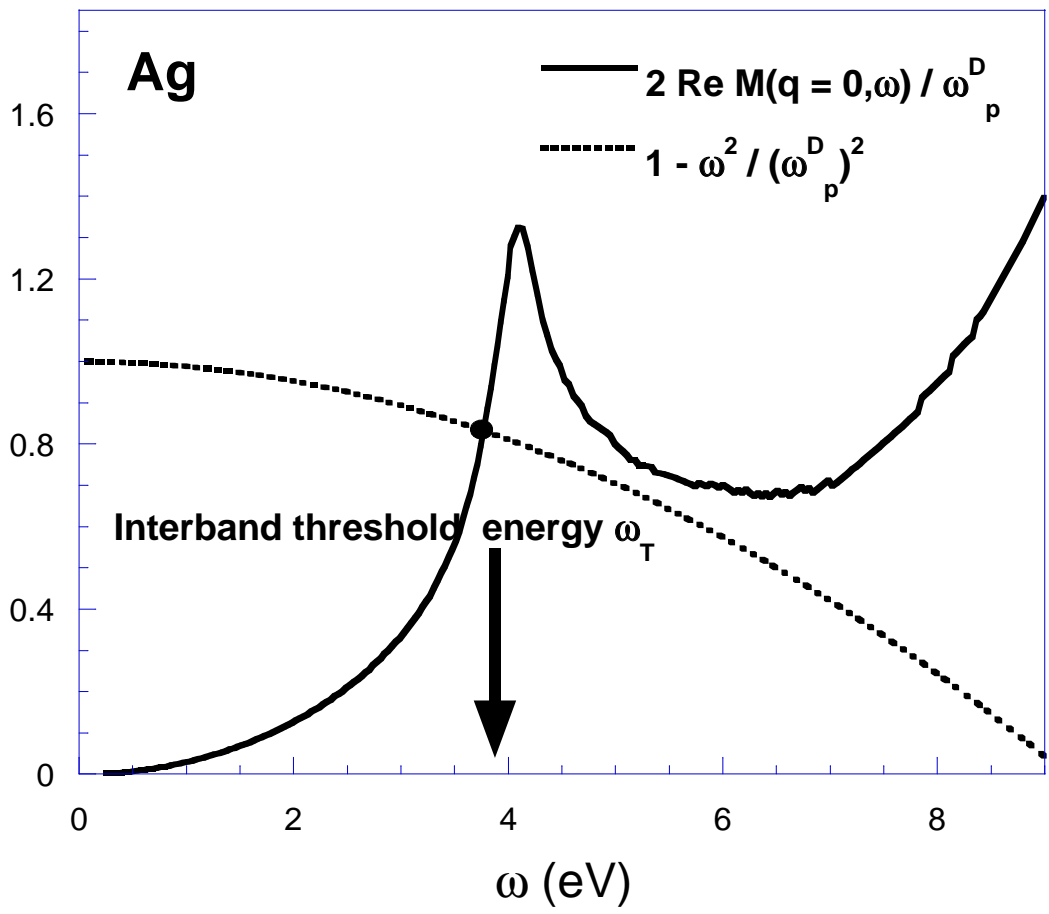
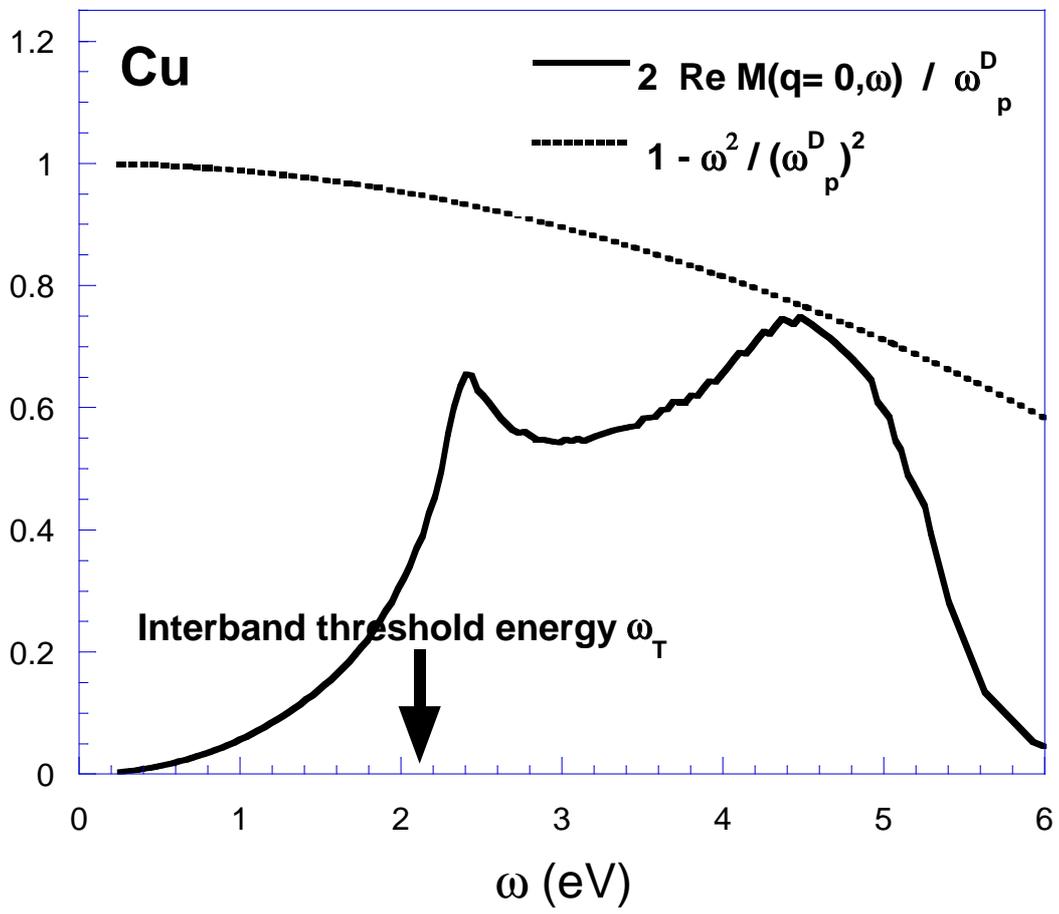
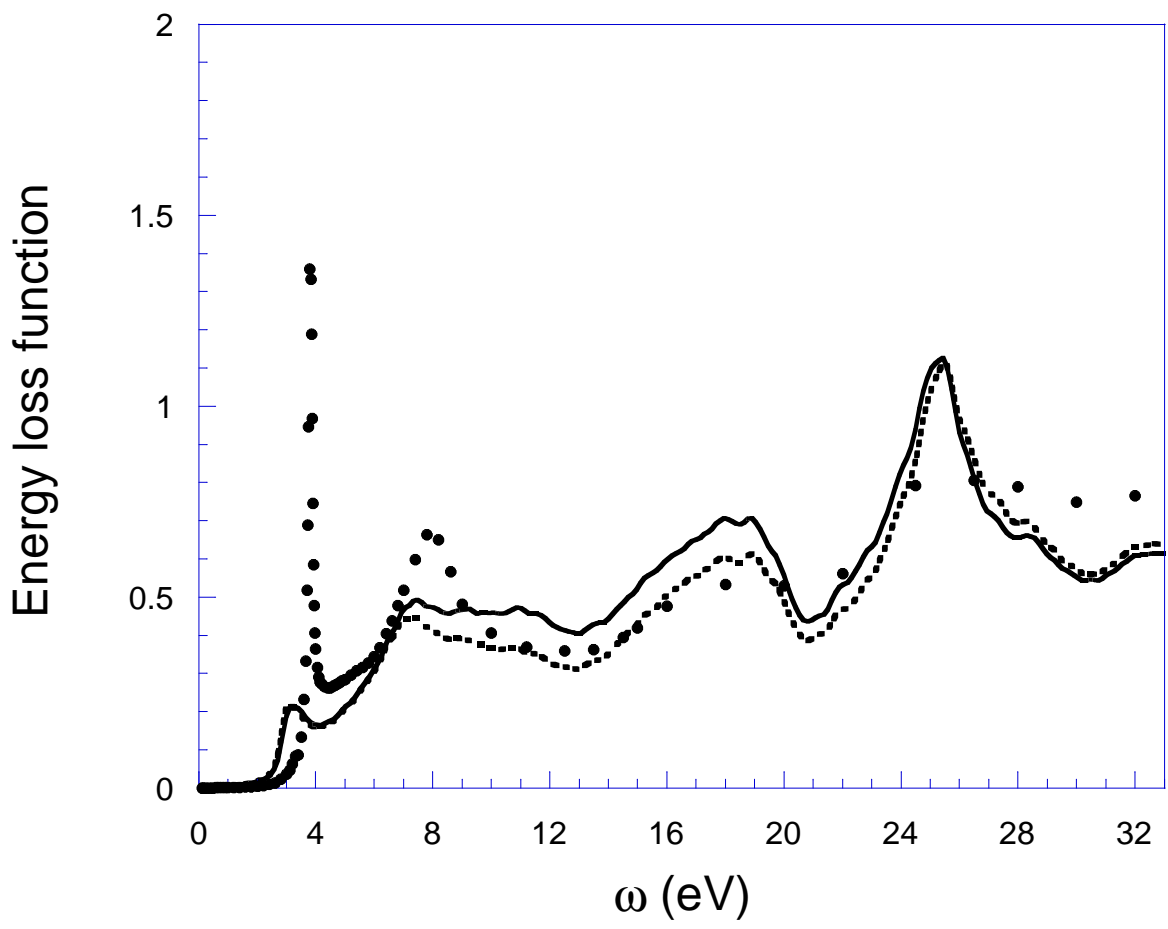
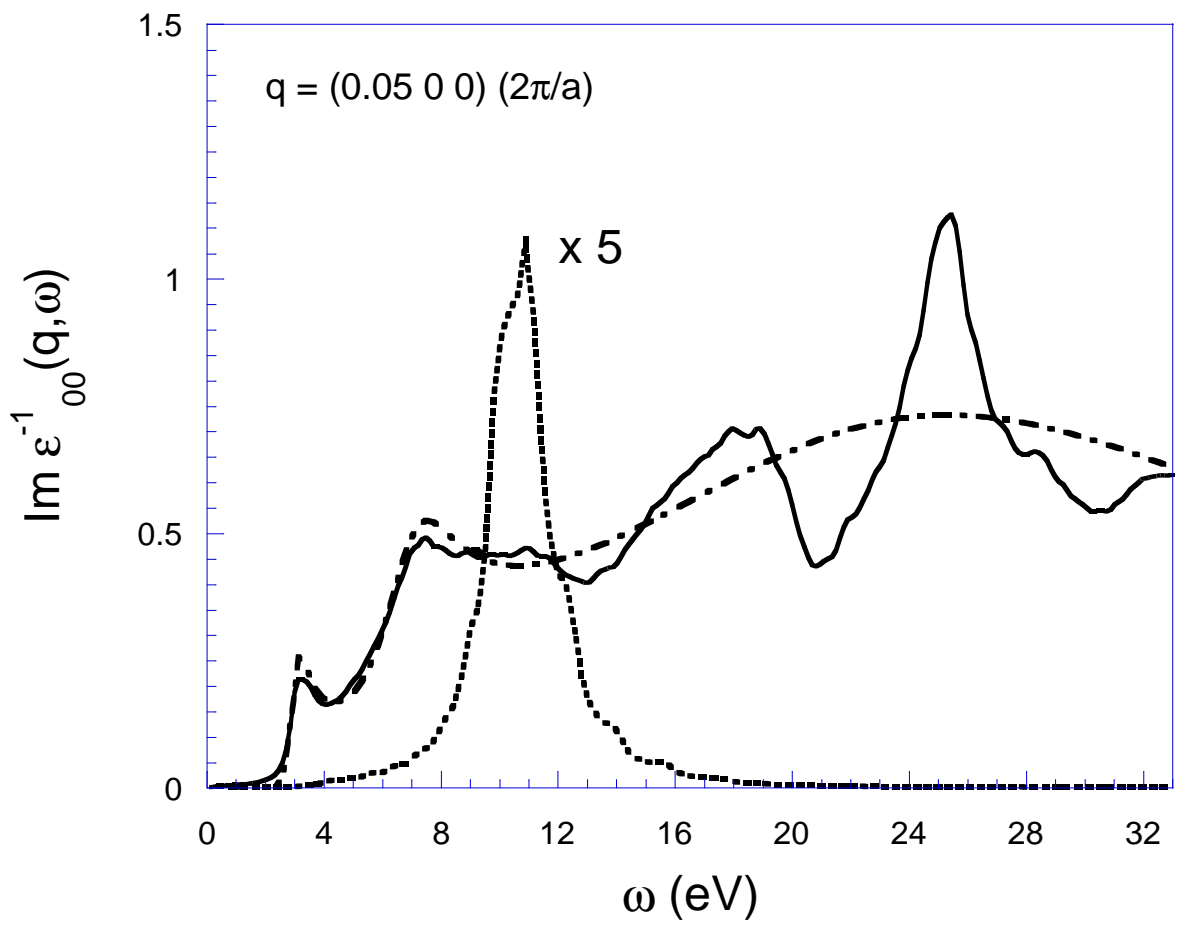
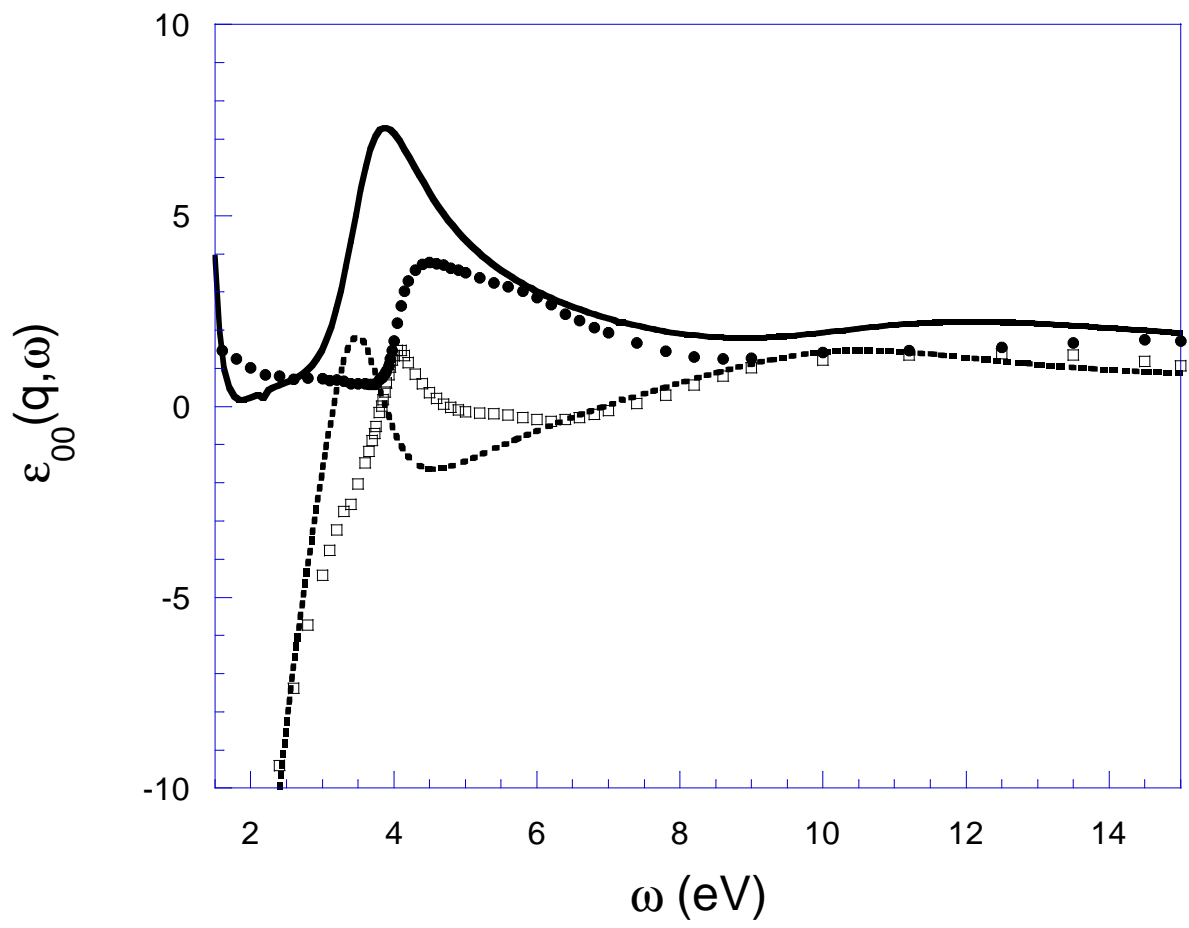
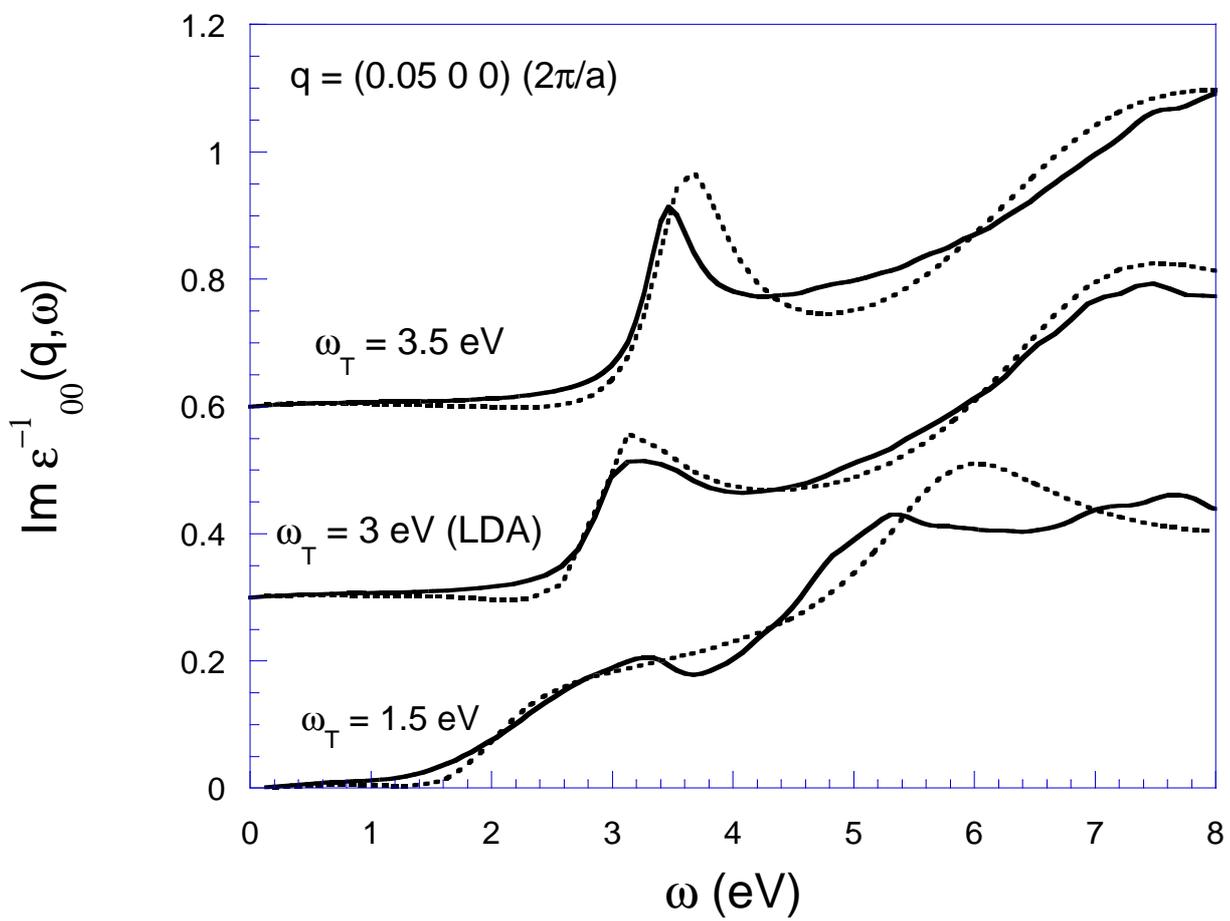
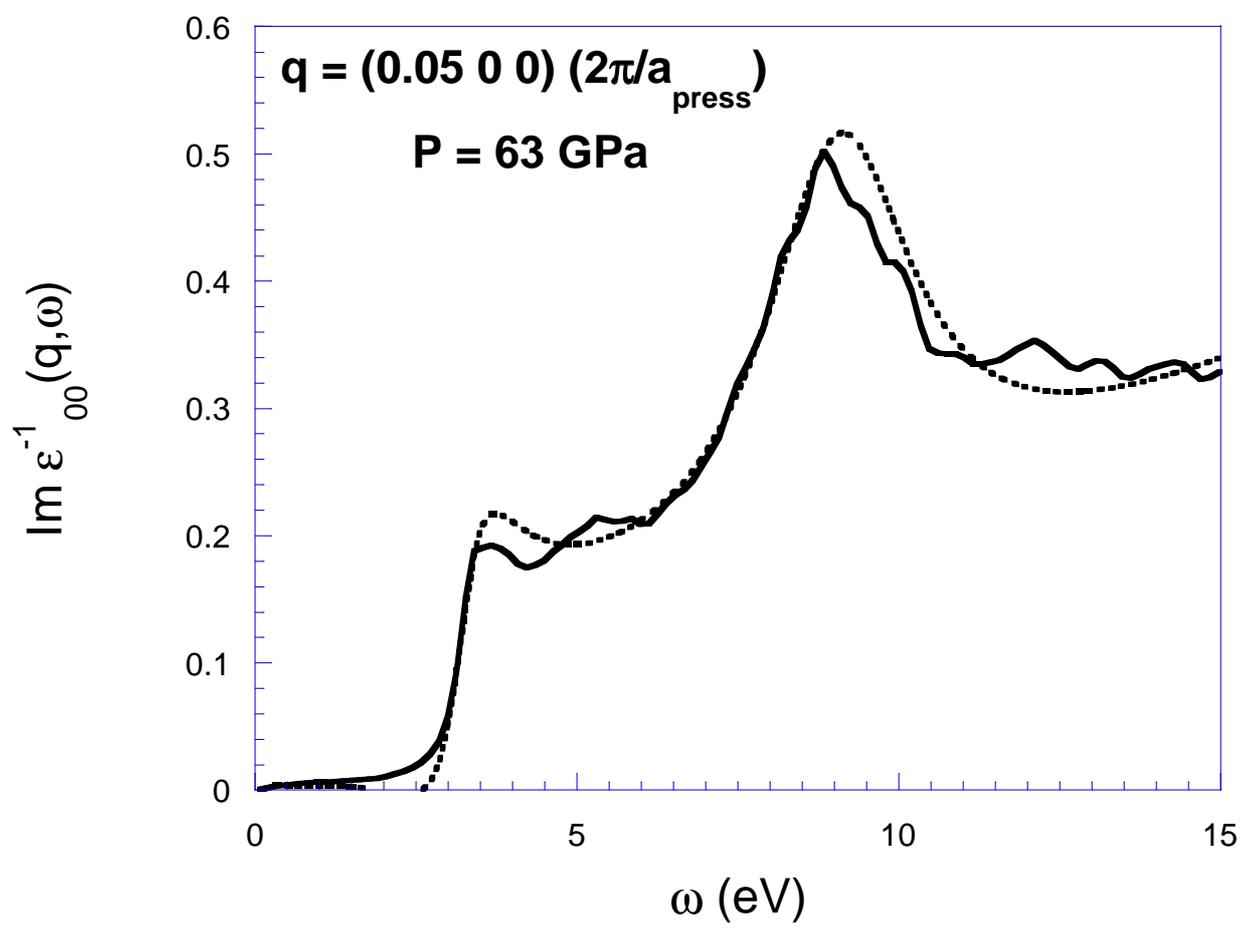
\end{document}